\documentclass[onecolumn]{revtex4}
\usepackage{amsmath,amssymb,graphics,epsfig,subfigure}
\usepackage{color}
\usepackage[colorlinks,linkcolor=red,anchorcolor=red,citecolor=green]{hyperref}
\usepackage{setspace}
\usepackage{booktabs}
\usepackage{amsfonts}
\usepackage{bbm}
\usepackage{graphicx}
\usepackage{pgfplots}
\usepackage{mathrsfs}
\usepackage{color}
\usepackage{verbatim}
\usepackage{cancel}
\usepackage{float}
\begin{document}
	\thispagestyle{empty}
\begin{center}
	\title{Thermodynamic Supercritical Scaling Law for the charged Gauss-Bonnet AdS Black Hole}
	
	\author{Jia-Yi Xia$^{1}$ and Zhen-Ming Xu$^{1,2,3,4}$
	\vspace{6pt}\\}

\affiliation{$^{1}$School of Physics, Northwest University, Xi'an 710127, China\\
	$^{2}$Peng Huanwu Center for Fundamental Theory, Xi'an 710127, China\\
	$^{3}$Shaanxi Key Laboratory for Theoretical Physics Frontiers, Xi'an 710127, China\\
	$^{4}$Fundamental Discipline Research Center for Quantum Science and technology of Shaanxi Province, China}
	
	\begin{abstract}
The phase structure of the charged Gauss-Bonnet AdS black holes varies with the spacetime dimension. Within an appropriate parameter range, only the six-dimensional case admits two physical critical points and it is an ideal model for investigating supercritical thermodynamics in black hole systems with multiple critical points. In this study, we investigate the supercritical thermodynamic properties of this model in the extended phase space and construct the supercritical phase diagrams associated with the two critical points. The thermodynamic supercritical scaling analysis for the supercritical crossover lines in the vicinity of both critical points is completed, demonstrating universality. The corresponding coefficients for the external field scaling and the order parameter scaling are obtained, and their physical significance is discussed.
	\end{abstract}

\maketitle
\end{center}

\section{Introduction}

In classical thermodynamics, the liquid-gas phase transition of a single-component fluid is a typical first-order phase transition. When the temperature and pressure are below their respective critical values, the liquid and gas phases can coexist under phase equilibrium conditions, separated by the coexistence curve of the first-order phase transition~\cite{2104.10619}. As the temperature and pressure approach the critical point from below, the liquid and gas phases are becoming increasingly indistinguishable, and the coexistence curve terminates at the critical point. At the critical point, the system undergoes a continuous phase transition. Upon further increasing the temperature and pressure beyond the critical point, the system enters the supercritical state, where no well-defined phase boundary exists in the strict sense; different thermodynamic states can interconvert continuously without crossing a first-order phase transition.

Recently, the thermodynamic behavior in the supercritical region has emerged as a new research direction in black hole thermodynamics, providing a novel perspective for characterizing the phase structure of black holes beyond the critical point and deepening the thermodynamic analogy between gravitational systems and conventional fluids~\cite{2505.24148,2504.05708,2504.04995}.

In anti-de Sitter (AdS) spacetime, it was proposed that the cosmological constant can be interpreted as thermodynamic pressure~\cite{0904.2765,1608.06147}. This proposal established the framework of black hole thermodynamics in the extended phase space and demonstrated that the charged black hole undergoes a phase transition between small-black-hole (SBH) and large-black-hole (LBH), remarkably analogous to that of the van der Waals fluid~\cite{1205.0559}. Since then, extensive studies have been devoted to black hole phase structures, critical phenomena, critical exponents, and universality classes \cite{1208.6251,1209.1707}.

Previous studies of black hole critical phenomena have mainly focused on the first-order phase transition below the critical point. Above the critical point, namely in the supercritical regime, however, the distinction between small and large black holes gradually disappears. In~\cite{2504.05708}, using complex analysis, the supercritical crossover of the Reissner-Nordstr\"{o}m-AdS (RN-AdS) were obtained, demonstrating the thermodynamic supercritical behavior of the black hole thermodynamic system. While in~\cite{2506.10808}, a refined supercritical phase diagram was constructed for this black hole. By extending the critical isochore into the supercritical region and identifying the loci of extrema of the isothermal compressibility, two supercritical crossover lines were defined, which divide the supercritical region into an SBH-like region, an LBH-like region, and an SBH-LBH-indistinguishable region between the two phases. Moreover, it was shown that both crossover lines obey universal power-law scaling relations in the vicinity of the critical point.

As the second-order term of Lovelock gravity \cite{lovelock}, Gauss-Bonnet gravity is the unique nontrivial extension of Einstein gravity that preserves second-order field equations in spacetime dimensions $D\geq5$ \cite{10.1103,0109133}. It has therefore become one of the most representative higher-curvature correction models in higher-dimensional gravity theories. More importantly, it is not only a natural consequence of the low-energy effective theory of string theory, but also provides a fundamental theoretical framework for investigating quantum gravity corrections \cite{156.315}. Previous studies have shown that the Gauss-Bonnet coupling significantly modifies the thermodynamic equation of state of black holes. It not only shifts the location of the critical point and the corresponding critical parameters, but also gives rise to rich phase structures that are absent in Einstein gravity, including reentrant phase transitions, multiple critical points, and triple points \cite{1406.7015,1306.6233,1412.5028}.

The emergence of multiple critical points raises an important question that has not yet been addressed in previous studies of black hole supercritical thermodynamics, namely whether the supercritical thermodynamics established for a black hole with a single critical point can be extended to that with multiple critical points. To investigate this issue, in this paper we consider the six-dimensional charged Gauss-Bonnet AdS (GB-AdS) black hole, which exhibits two physical critical points within an appropriate parameter range. Against this background, we further validate the universal scaling law of supercritical phenomena, in order to enrich the research on the thermodynamic refinement of supercritical phenomena in black holes for understanding the microscopic thermodynamic nature of black holes with higher-curvature corrections.

In Sec.~\ref{II}, we briefly review the thermodynamic properties of six-dimensional charged GB-AdS black holes and calculate several thermodynamic quantities required in this study. In Sec.~\ref{3}, we introduce the supercritical crossover lines and present the supercritical phase diagrams associated with the two critical points of the black hole. Based on these results, the supercritical scaling behavior in the vicinity of the two critical points is investigated in Sec.~\ref{V}. There is a summary in Sec.~\ref{Vv}.

\section{Thermodynamic properties of 6d charged Gauss-Bonnet AdS black hole}\label{II}
In six-dimensional spacetime, the action of Einstein-Gauss-Bonnet gravity is given by
\begin{equation}
S = \frac{1}{16\pi} \int d^{6}x \, \sqrt{-g} \,
\Bigl[
  R - 2\Lambda
  + \alpha_{\mathrm{GB}} \bigl( R_{\mu\nu\gamma\delta} R^{\mu\nu\gamma\delta} - 4 R_{\mu\nu} R^{\mu\nu} + R^{2} \bigr) - 4\pi F_{\mu\nu} F^{\mu\nu}
\Bigr],
\label{eq:action}
\end{equation}
where $\Lambda$ is the cosmological constant; $\alpha_{\mathrm{GB}}$ is the Gauss-Bonnet coupling coefficient with dimension $(\mathrm{length})^{2}$ ; the Maxwell field strength is $F_{\mu\nu} = \partial_\mu A_\nu - \partial_\nu A_\mu$, where $A_\mu$ is the vector potential. The metric form of the static black hole solution is
\begin{equation}
ds^2 = -f(r)\,dt^2 + f^{-1}(r)\,dr^2 + r^2\left[ d\theta_1^2 + \sin^2\theta_1\,d\theta_2^2 + \cos^2\theta_1\left(d\theta_3^2 + \sin^2\theta_3\,d\theta_4^2\right)\right],
\end{equation}
where the $f(r)$ function is~\cite{1306.6233,1402.2837}
\begin{equation}
f(r) = 1 + \frac{r^{2}}{2\alpha_{\mathrm{eff}}}
\left[
  1 - \sqrt{
    1 + \frac{\alpha_{\mathrm{eff}}}{2}
    \left(
      \frac{32\pi M}{\Sigma_{4}\, r^{5}}
      - \frac{4Q^{2}}{3r^{8}}
      + \frac{4\Lambda}{5}
    \right)
  }
\right],
\end{equation}
here the effective Gauss-Bonnet coefficient $\alpha_{\mathrm{eff}} = 6\alpha_{\mathrm{GB}}$, and $\Sigma_4$ is the area of a unit 4-dimensional sphere. The Gauss-Bonnet coefficient is usually taken to be positive in the string-theory context, while the possibility of a negative coupling and its discussion can also refer the literature~\cite{2208.13360,2103.00257}. In the extended phase space, the cosmological constant is regarded as pressure via $P=-\Lambda/(8\pi)$, resulting that the equation of state for the black hole is
\begin{equation}
P = T \frac{v^{2} + 2}{v^{3}} - \frac{6v^{6} + 2v^{4} - Q^{2}}{8\pi v^{8}},
\label{eq7}
\end{equation}
where $T$ is the Hawking temperature, the specific volume is given by $v = 4\,r_h/(d - 2) = r_h$ and $r_h$ is the event horizon radius of the black hole.

It was shown in Ref.~\cite{1402.2837} that, for $d=6$, the charged Gauss-Bonnet black hole exhibits a triple point and a coexistence of small, intermediate, and large black holes when $0.1705<Q<0.1946$ (details are provided in Appendix~\ref{A}). Hence in this paper, we set $\Sigma_4=1$, $\alpha_{\mathrm{eff}}=1$, and $Q = 0.18$, which lies within this parameter regime. As long as the system remains within the parameter regime where multiple physical critical points exist, we expect the overall qualitative features of the supercritical behavior to remain unchanged.

The critical point is defined as the inflection point of the isotherm, satisfying the constraint conditions where the first and second isothermal partial derivatives vanish simultaneously
\begin{equation}
\left( \frac{\partial P}{\partial v} \right)_T = \left( \frac{\partial^2 P}{\partial v^2} \right)_T = 0.
\label{eq8}
\end{equation}
Thus, mathematically, three critical points can be solved from~(\ref{eq8}), which are respectively
\begin{equation}
\begin{aligned}
T_{c1} &= 0.112381, & P_{c1} &= 0.0197442, & v_{c1} &= 1.55453, \\
T_{c2} &= 0.112084, & P_{c2} &= 0.0192661, & v_{c2} &= 1.13426, \\
T_{c3} &= 0.112711, & P_{c3} &= 0.021548,  & v_{c3} &= 0.792783.
\end{aligned}
\end{equation}
Among them, although $(T_{c2}, P_{c2}, v_{c2})$ satisfies Eq.~(\ref{eq8}) mathematically, the literature~\cite{1402.2837} showed that it does not correspond to the endpoint of any two-phase coexistence line and therefore is not regarded as a physical critical point. While $(T_{c1}, P_{c1}, v_{c1})$ corresponds to the IBH-LBH phase transition endpoint, and $(T_{c3}, P_{c3}, v_{c3})$ corresponds to the SBH-IBH phase transition endpoint. The next section will focus on these two critical points to further discuss the thermodynamic behavior of the six-dimensional charged GB-AdS black hole in the supercritical region.
	   	
\section{Supercritical Crossover Lines of 6d Charged Gauss-Bonnet AdS Black Hole}\label{3}

In the supercritical region, although no genuine phase boundary exists, thermodynamic crossovers between different states can still be characterized by several preliminary methods, including the projection of complex zeros on the real axis~\cite{2504.05708}, the higher-order derivatives of the free energy~\cite{2504.04995}, the dynamical definition~\cite{2505.24148} and the extrema of appropriate thermodynamic response functions~\cite{2506.10808,2304.04244}. In this paper, in order to refine the supercritical crossover, we adopted the supercritical analysis method developed in~\cite{2304.04244}, which was further extended to black hole thermodynamic systems in~\cite{2506.10808}.

We adopt the isothermal compressibility as the response function for six-dimensional charged GB-AdS black hole
\begin{equation}
\kappa_T = -\frac{1}{v} \frac{\partial v}{\partial P}=\frac{2\pi v^8}{2\pi T v^7 + 12\pi T v^5 - 3v^6 - 2v^4 + 2Q^2}.
\end{equation}

To construct the supercritical crossover lines, we first extend the critical isochore into the supercritical region and construct a series of paths parallel to it. Along each path, the isothermal compressibility $\kappa_T$ is evaluated, and the point at which $\kappa_T$ reaches its maximum is identified. Repeating this procedure for paths on both sides of the critical isochore yields two sets of extremal points. These points collectively define the two supercritical crossover lines $L_+$ and $L_-$, located on the two sides of the critical isochore \cite{2506.10808,2304.04244}.

\begin{figure}[H]
    \centering
    \begin{minipage}[t]{0.43\linewidth}
        \centering
        \includegraphics[width=\textwidth]{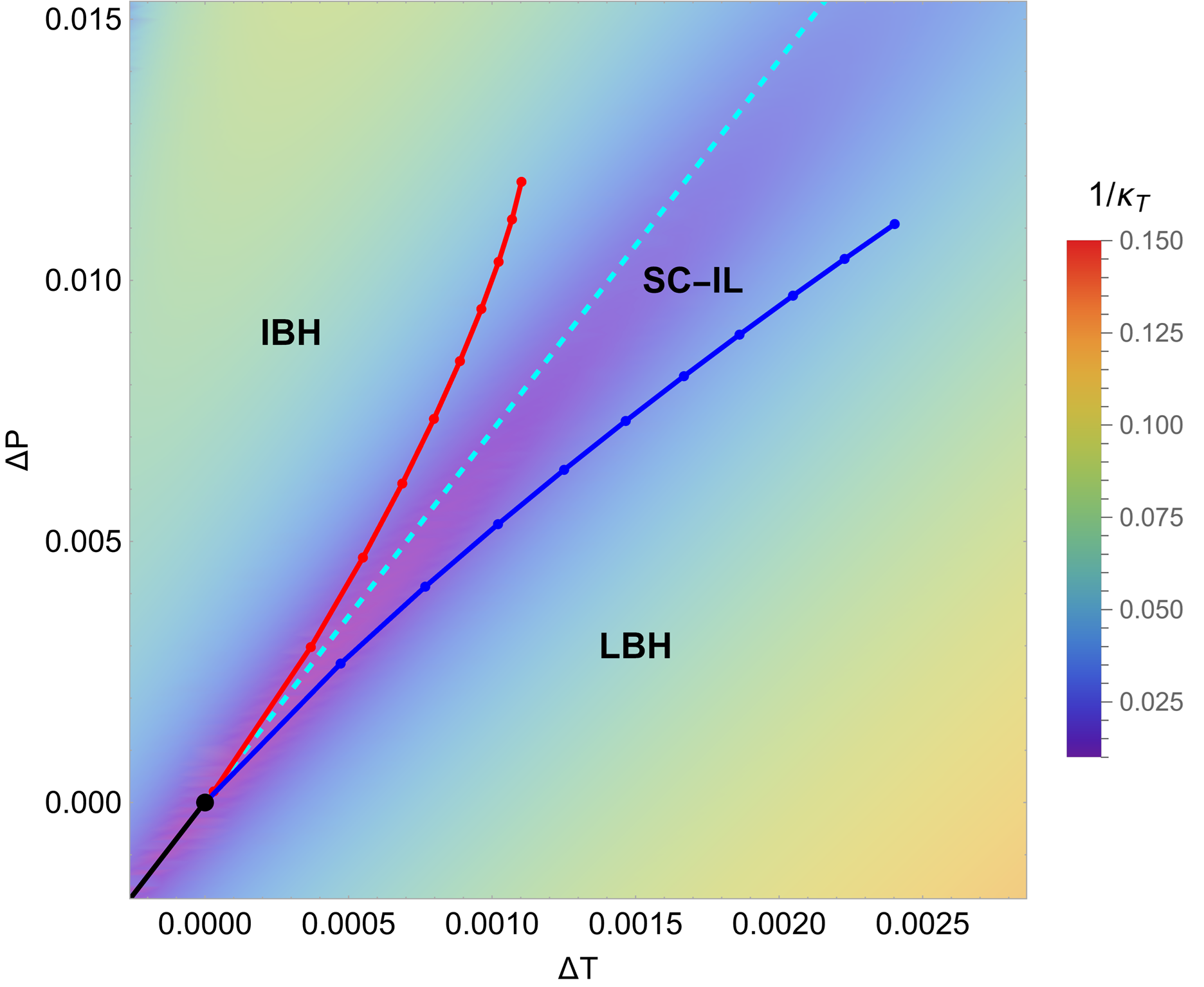}
        \small\textbf{(A)}
    \end{minipage}\hfill
    \begin{minipage}[t]{0.45\linewidth}
        \centering
        \includegraphics[width=\textwidth]{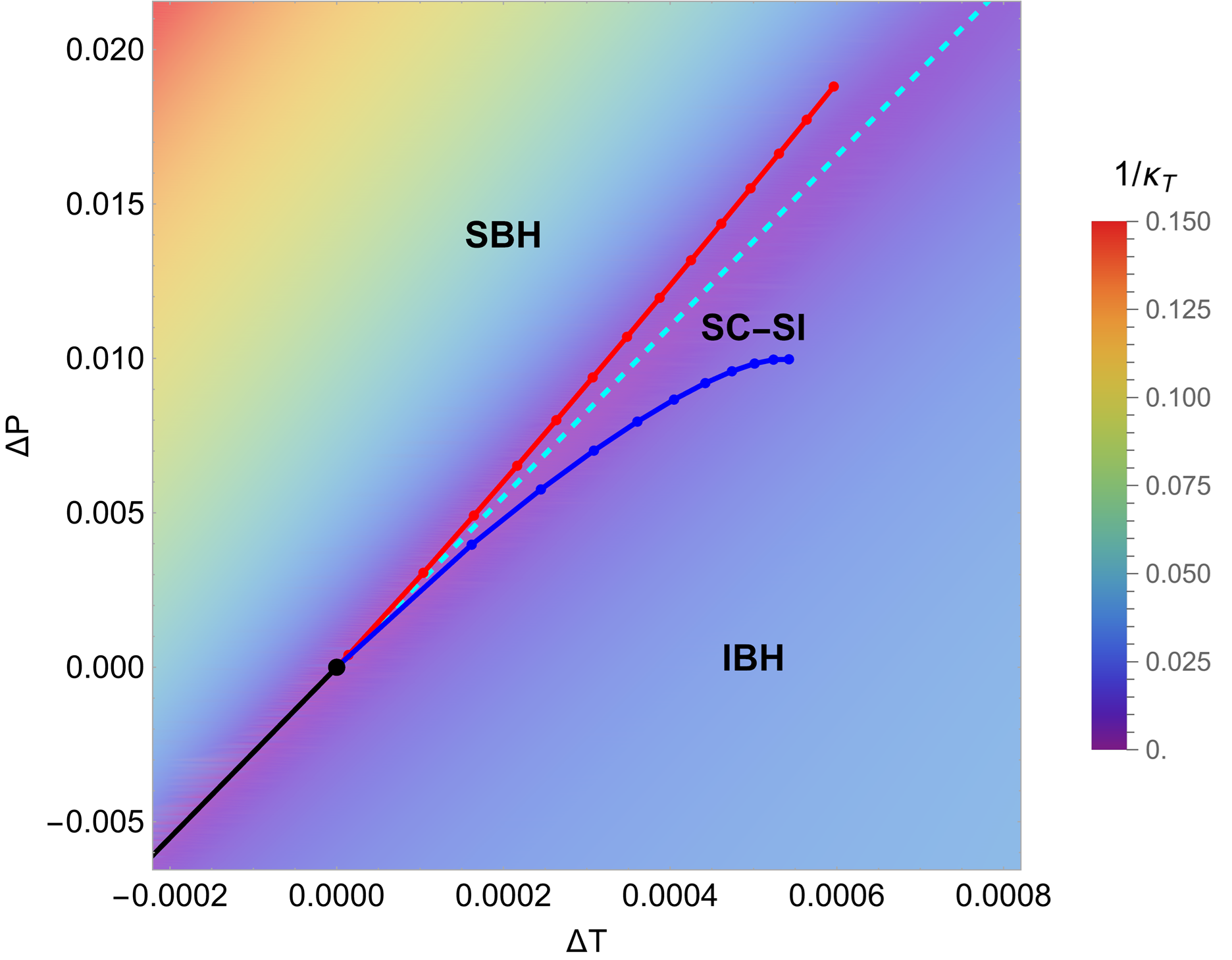}
        \small\textbf{(B)}
    \end{minipage}
    \caption{Supercritical Phase diagram at the endpoint of (A) the IBH--LBH phase transition (the $L_{\pm}$ lines divide the black hole phase diagram into three regions: IBH, IBH-LBH-indistinguishable, and LBH), (B) the SBH--IBH phase transition (the $L_{\pm}$ lines divide the black hole phase diagram into three regions: SBH, SBH-IBH-indistinguishable, and IBH). The data have been normalized, and both the horizontal and vertical axes are rescaled by the corresponding critical values: $\Delta P = P/P_c - 1$ and $\Delta T = T/T_c - 1$. The background color is determined by the magnitude of the inverse isothermal compressibility, $1/\kappa_T$. The values shown in the color bar are intended only to illustrate the relative variation of $1/\kappa_T$ and do not represent its actual physical values. Black solid dots represent two critical points.}
    \label{L}
\end{figure}

Based on this spirit, we respectively present the $L_{\pm}$ lines at the two critical points of the six-dimensional charged GB-AdS black hole, which are shown in Fig.~\ref{L}. These lines are qualitatively consistent with those obtained for the four-dimensional charged AdS black hole in~\cite{2506.10808}. The result indicates that, although the six-dimensional charged GB-AdS black hole possesses a much more complicated phase structure than its four-dimensional charged AdS counterpart, its supercritical behavior in the vicinity of the critical point still exhibits similar geometric features and evolution patterns. In the vicinity of the critical isochore, $1/\kappa_T$ reaches a local minimum and gradually approaches zero as the critical point is approached, corresponding to a continuous increase in the isothermal compressibility, which eventually diverges at the critical point.

\section{Universal scaling laws of 6d charged Gauss-Bonnet AdS black hole}\label{V}
One of the important features of critical phenomena is their universality, meaning that although different physical systems have different microscopic structures and interaction mechanisms, they often exhibit similar critical behavior near the critical point.

After the $L_{\pm}$ lines were proposed in~\cite{2304.04244}, the following scaling relations were found in several specific black hole models~\cite{2407.05455}. Near the critical temperature $T_c$ , there is
\begin{equation}
\delta P^{\pm} \propto (T - T_c)^{\beta + \gamma} \quad \text{and}\quad
\delta \rho^{\pm} \propto (T - T_c)^{\beta},
\end{equation}
where $T$ is considered as the control field, $P$ is considered as the conjugate external field, $\rho$ is the order parameter, $\beta$ and $\gamma$ are scritical exponents, whose specific values depend on the universality class.

For six-dimensional charged GB-AdS black holes in the extended phase space, the Hawking temperature $T$ serves as the control field, the pressure $P$ corresponds to the external field, and the black hole number density $\rho = 1/v$~\cite{1502.00386} is chosen as the order parameter. In order to study the scaling behavior of the external field and order parameter relative to their critical values near the supercritical crossover lines, we define $\delta P = P(v, T) - P(v_c, T)$ and $\delta\rho = \rho - \rho_c$. In reference~\cite{2506.10808}, the authors calculated the supercritical scaling laws of some black hole thermodynamic systems with a single critical point and provided the mean-field universality class, with critical exponents $\beta = 1/2$ and $\gamma = 1$. For the case of multiple critical points now, we therefore take the mean-field predictions as a reference and examine whether the numerical scaling behavior of the $L_{\pm}$ lines is consistent with them.

\begin{figure}[H]
    \centering
    \begin{minipage}[t]{0.38\linewidth}
        \centering
        \includegraphics[width=\textwidth]{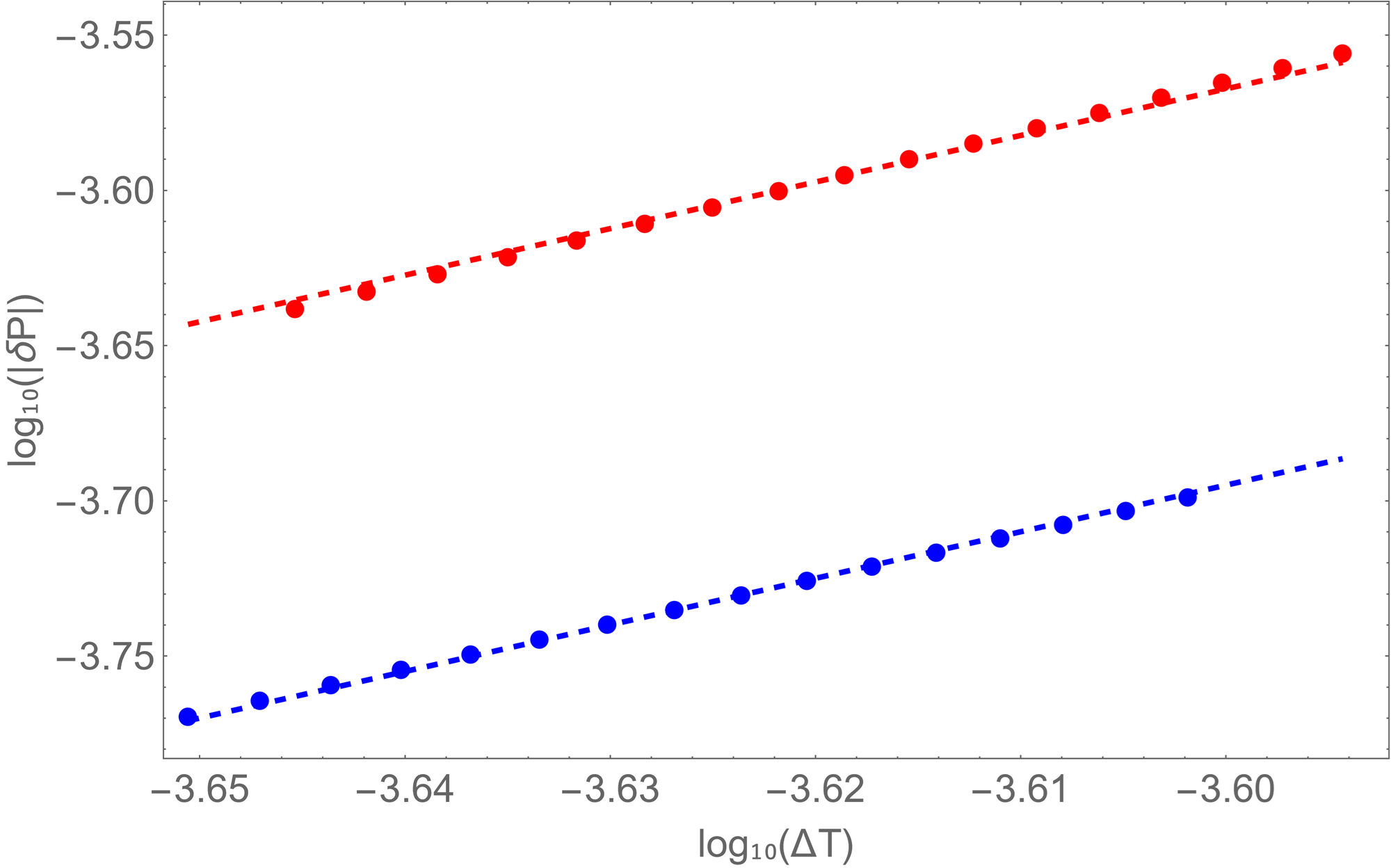}
    \end{minipage}\hfill
    \begin{minipage}[t]{0.51\linewidth}
        \centering
        \includegraphics[width=\textwidth]{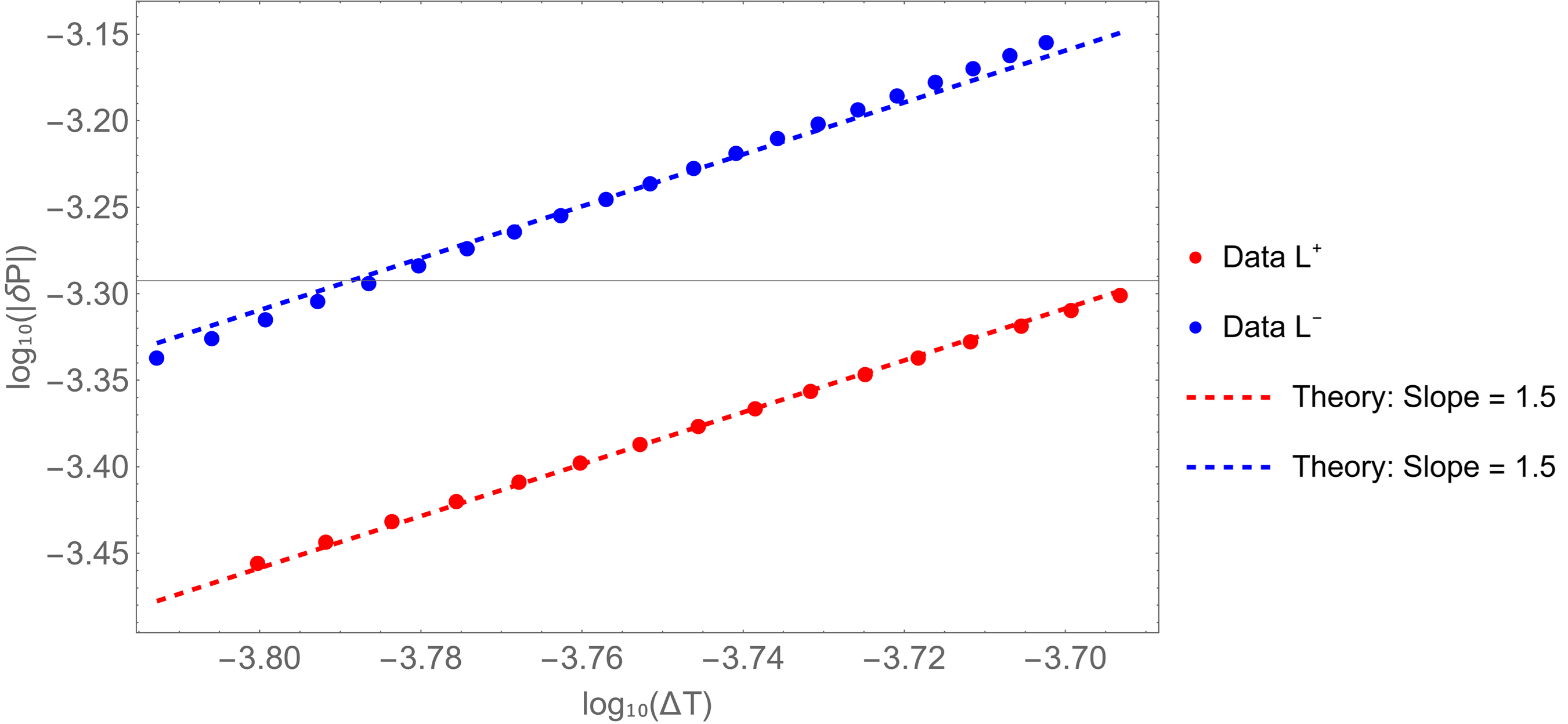}
    \end{minipage}
    \caption{Scaling law of the external field. Left panel: at the IBH-LBH phase transition endpoint; Right panel: at the SBH-IBH phase transition endpoint (both horizontal and vertical axes are plotted on a logarithmic scale).}
    \label{alpha}
\end{figure}

\begin{figure}[H]
    \centering
    \begin{minipage}[t]{0.4\linewidth}
        \centering
        \includegraphics[width=\textwidth]{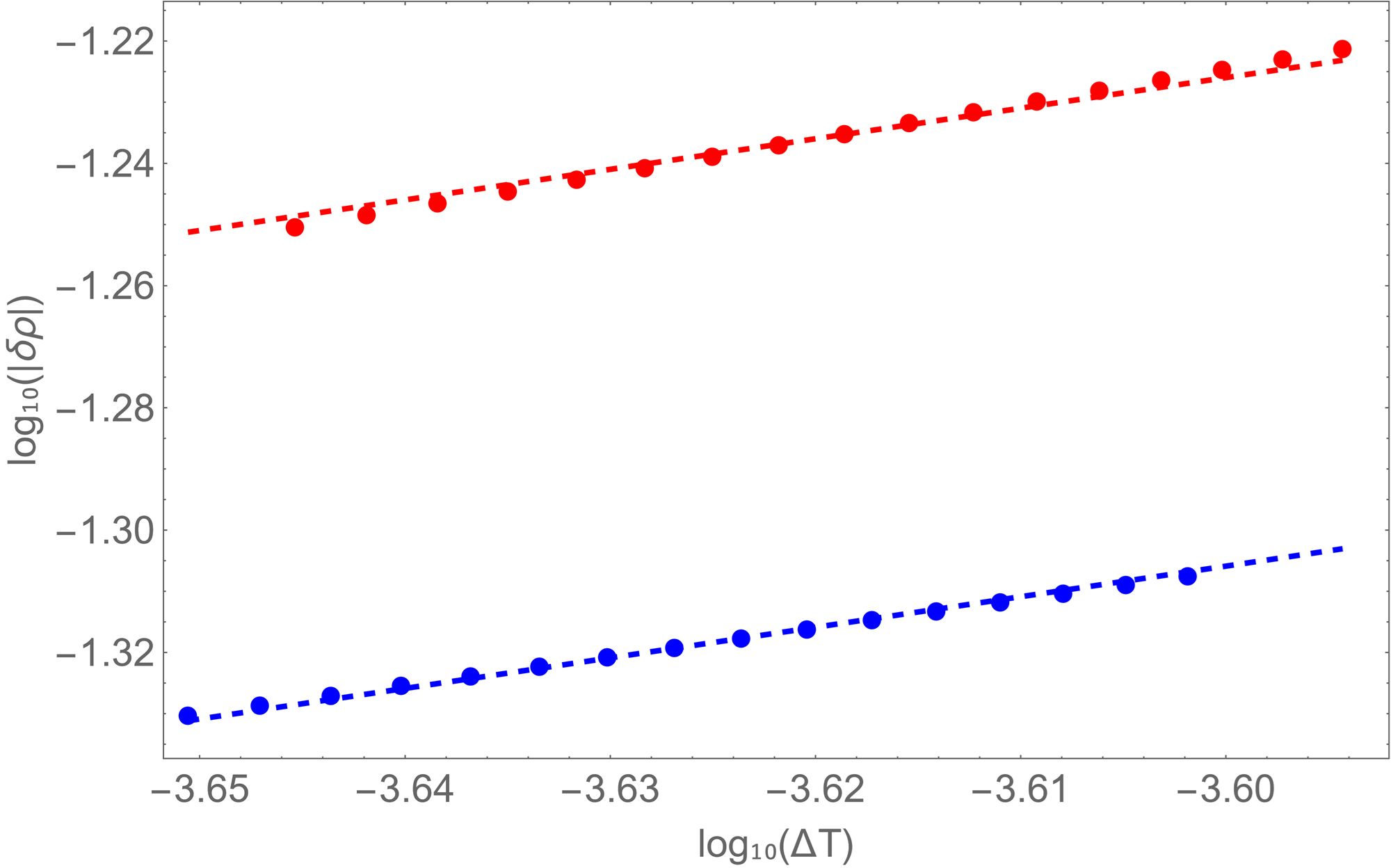}
    \end{minipage}\hfill
    \begin{minipage}[t]{0.51\linewidth}
        \centering
        \includegraphics[width=\textwidth]{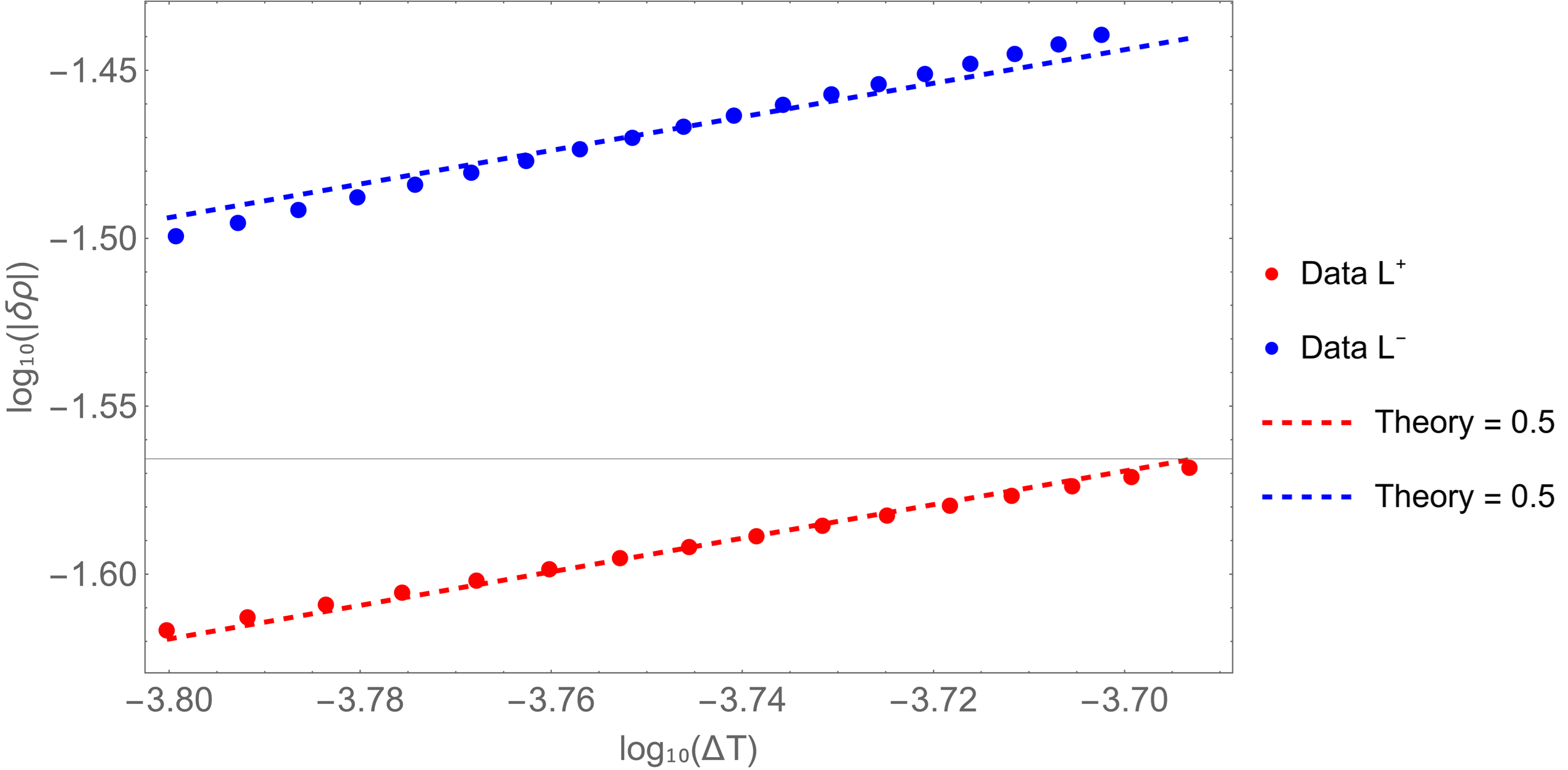}
    \end{minipage}
    \caption{Scaling law of the order parameter. Left panel: at the IBH-LBH phase transition endpoint; Right panel: at the SBH-IBH phase transition endpoint.}
    \label{beta}
\end{figure}

Accordingly, we adopt $\alpha = \beta + \gamma = 3/2$ and $\beta = 1/2$ as the theoretical slopes for the external-field and order-parameter scaling, respectively. Based on these theoretical slopes, we construct the corresponding reference lines and calculate the nonuniversal scaling coefficients $\alpha_{\pm}$ and $\beta_{\pm}$ at the two critical points, which are shown in Figs.~\ref{alpha},~\ref{beta} and Table~\ref{table1}.

\begin{table}[H]
    \centering
    \caption{Values of the coefficients $\alpha_{\pm}$ and $\beta_{\pm}$.}
    \label{table1}

    \begin{tabular}{lcccc}
        \hline \hline
         & $\alpha_+$ & $\alpha_-$ & $\beta_+$ & $\beta_-$ \\
        \hline
        IBH-LBH & 68.027  & 50.704  & 3.750  & 3.120  \\
        $\sigma$ & 0.0018 & 0.0008 & 0.0011 & 0.0005 \\
        \hline
        SBH-IBH & 174.372 & 245.783 & 1.908 & 2.548 \\
        $\sigma$ & 0.0017 & 0.0051 & 0.0016 & 0.0035 \\
        \hline \hline
    \end{tabular}
\end{table}

Figure~\ref{alpha} shows the scaling behavior of the external field $\delta P^{\pm}$ near the two critical points, with theoretical values and datasets. The reference straight lines in both panels have the theoretical slope $\alpha=3/2$.
The numerical data for both crossover lines near the two critical points are in good agreement with the reference lines. This indicates that the external-field scaling behavior approximately follows the predictions of mean-field theory near both critical points.

Figure~\ref{beta} shows the scaling behavior of the order parameter $\delta \rho^{\pm}$ near the two critical points and the reference straight lines in both panels have the theoretical slope $\beta=1/2$.
The numerical data for the order parameter near both critical points are in good agreement with the reference lines, which also indicates that the order-parameter scaling behavior approximately follows the predictions of mean-field theory near both critical points.

Table~\ref{table1} summarizes the fitted coefficients $\alpha_{\pm}$ and $\beta_{\pm}$, together with the corresponding root mean square errors $\sigma$. $\alpha_{\pm}$ reflects the speed at which the $L_{\pm}$ lines deviate from the critical isochore; under the same $\Delta T$ , the larger the value of $\alpha_{\pm}$ , the larger the corresponding $\Delta P$ , and the farther the $L_{\pm}$ lines are from the critical isochore. On the other hand, $\beta_{\pm}$ reflects the amplitude of the order parameter's deviation from the critical value along the $L_{\pm}$ lines; under the same $\Delta T$ , the larger the value of $\beta_{\pm}$ , the more obvious the deviation of $1/v$ from the critical value. Unlike $\alpha$ and $\beta$, the values of $\alpha_{\pm}$ and $\beta_{\pm}$ depend on the specific model and thermodynamic path and do not possess universality. The $\sigma$ quantifies the deviations of the numerical data from the reference scaling lines. The relatively small values of $\sigma$ indicate that the numerical results at both critical points are in good agreement with the reference mean-field scaling law. The data in Table~\ref{table1} indicates that our current numerical conclusion is reliable.
	   	
\section{Conclusion}\label{Vv}
In this paper, we investigate the thermodynamic properties of the supercritical region of six-dimensional charged GB-AdS black holes within the framework of extended phase space thermodynamics. Considering the characteristic feature that this model possesses two critical points, we construct the supercritical phase diagrams in the vicinity of the two critical points separately and calculate the universal scaling laws of the supercritical crossover lines.

The results show that the overall structures of the supercritical crossover lines near the two critical points exhibit similar features to those of the four-dimensional charged AdS black hole, indicating that the supercritical crossover line, as an important tool for characterizing continuous thermodynamic changes in the supercritical region, is also applicable to higher-dimensional Gauss-Bonnet black holes with rich phase structures. The scaling behavior in the supercritical region is found to be consistent with mean-field theory predictions in~\cite{2506.10808}.

In the future, the supercritical behavior of black holes in different gravitational theories, spacetime dimensions, and other models with rich phase structures can be further investigated. Furthermore, existing analytical studies~\cite{2509.17038} have demonstrated that an emergent $\mathbb{Z}_2$ symmetry arises near the critical point of the four-dimensional charged AdS black holes. As shown in Table~\ref{table1}, the six-dimensional charged Gauss–Bonnet AdS black hole appears to exhibit a breakdown of this symmetry. However, this numerical discrepancy cannot serve as direct evidence for such a symmetry breakdown in the critical scaling regime. Whether the symmetry breaking truly exists and its underlying influencing factors remain open questions for future investigation.

\section*{Acknowledgments}
We thank professor Yen-Chin Ong for suggesting valuable comments and suggestions on our manuscript and Zhi-Yuan Li for useful discussions. This research was supported in part by the National Natural Science Foundation of China (Grant Nos. 12575064, 12247103), by the Natural Science Basic Research Plan in Shaanxi Province of China (Grant No. 2025JC-YBQN-029), and by the project of Tang Scholar in Northwest University.

\appendix
\section{Critical points of the $d$-dimensional GB-AdS black holes}
\label{A}
We start with the equation of state of the $d$-dimensional GB-AdS black holes. The effective GB coefficient $\alpha_{\mathrm{eff}} = (d - 3)(d - 4)\alpha_{\mathrm{GB}} > 0$. The equation of state of the black hole for the spherical horizon is
\begin{equation}
    P = \frac{(d - 2)T}{4r_h} \left(1 + \frac{2\alpha_{\text{eff}}}{r_h^2}\right)
        - \frac{(d - 2)(d - 3)}{16\pi r_h^2}
        - \frac{(d - 2)(d - 5)\alpha_{\text{eff}}}{16\pi r_h^4}
        + \frac{Q^2}{8\pi r_h^{2d-4}}.
\label{1}
\end{equation}

The critical points satisfy the equation
\begin{equation}
    \left(\frac{\partial P}{\partial r_h}\right)_T = 0, \quad
    \left(\frac{\partial^2 P}{\partial r_h^2}\right)_T = 0.
\label{2}
\end{equation}
According to the first derivative of the equation of state with respect to the event horizon radius, we obtain the temperature along the spinodal curve
\begin{equation}
    T_{sp}(r) = \frac{(d - 3)r^2 + 2(d - 5)\alpha_{\text{eff}} - 2Q^2 r^{8-2d}}{2\pi r(r^2 + 6\alpha_{\text{eff}})}.
\end{equation}
While for the second derivative of the equation of state with respect to the event horizon radius, it is equivalent to the extremum condition of the spinodal temperature $dT_{sp}/dr = 0$. Hence we can obtain the central equation for the critical radius $r_c$
\begin{equation}
    0 = (d - 3)r_c^4 - 12\alpha_{\text{eff}} r_c^2 + 12(d - 5)\alpha_{\text{eff}}^2 - 2Q^2 r_c^{8-2d} \left[ (2d - 5)r_c^2 + 6(2d - 7)\alpha_{\text{eff}} \right].
\label{4}
\end{equation}
To facilitate the analysis of dimensional dependence, we introduce the dimensionless critical radius $u$ and dimensionless charge $h$:
\begin{equation}
    u = \frac{r_c^2}{\alpha_{\text{eff}}} > 0, \quad h = \frac{Q^2}{\alpha_{\text{eff}}^{d-3}} > 0,
\label{5}
\end{equation}
then the Eq.~(\ref{4}) can be rewritten in the dimensionless form
\begin{equation}
    F_d(u) = (d - 3)u^2 - 12u + 12(d - 5) - \frac{2h \left[ (2d - 5)u + 6(2d - 7) \right]}{u^{d-4}} = 0.
\label{6}
\end{equation}
The number of critical points at fixed charge corresponds exactly to the number of positive real roots of Eq.~(\ref{6}) for a given $h$. To identify the origin of the dimensional difference, we first examine the charge-independent quadratic polynomial in Eq.~(\ref{6})
\begin{equation}
    H_d(u) = (d - 3)u^2 - 12u + 12(d - 5).
\label{7}
\end{equation}
Its discriminant is
\begin{equation}
    \Delta_d = 144 - 48(d - 3)(d - 5) = -48(d - 2)(d - 6),
\label{8}
\end{equation}
and the sign of $\Delta_d$ is entirely determined by the spacetime dimension $d$, which governs the root structure of $H_d(u)$. For example, at $d=6$, we have $H_6(u) = 3u^2 - 12u + 12 = 3(u - 2)^2$ with a double root at $u=2$.

The discriminant alone is insufficient to determine the existence of multiple critical points. A complete analysis must incorporate the charge dependent term and count the number of solutions at fixed charge.

We now specialize to the $d=6$ case and the Eq.~(\ref{6}) becomes
\begin{equation}
    3(u - 2)^2 - \frac{2h(7u + 30)}{u^2} = 0,
\label{10}
\end{equation}
which can be rearranged into an explicit expression for the dimensionless charge as a function of $u$
\begin{equation}
    h_6(u) = \frac{3u^2(u - 2)^2}{2(7u + 30)}.
\label{11}
\end{equation}
The number of critical radii at fixed charge equals the number of intersections between the horizontal line $h = \text{constant}$ and the curve $h_6(u)$. Taking the derivative of $h_6(u)$ with respect to $u$ and setting the obtained expression to zero, we obtain two stationary points, $u = 2$ and $u = 20/21$. The point $u = 2$ is a local minimum with $h_6 = 0$, while $u = 20/21$ is a local maximum. The nonmonotonic behavior of $h_6(u)$ indicates that for a dimensionless charge between zero and this maximum value, a horizontal line at fixed $h$ intersects the curve at three distinct positive roots, corresponding to three positive algebraic critical radii rather than only two. Not all three algebraic solutions are physically meaningful, verification of the critical temperature $T_c$, critical pressure $P_c$ and Gibbs free energy is required to determine their thermodynamic relevance.

For completeness, we analyze the critical point structure in other dimensions. For $d = 5$, the dimensionless charge function reads
\begin{equation}
    h_5(u) = \frac{u^2(u - 6)}{5u + 18}.
\label{12}
\end{equation}
Positive charge requires $u > 6$, and $h_5(u)$ is strictly increasing over this domain, so there is exactly one positive critical radius for any positive charge.

For $d \ge 7$, the general form is
\begin{equation}
    h_d(u) = \frac{u^{d-4} \big[ (d-3)u^2 - 12u + 12(d-5) \big]}{2 \big[ (2d-5)u + 6(2d-7) \big]},
\label{13}
\end{equation}
where the quadratic factor in the numerator is strictly positive for $d\ge7$. Setting $n=d-4\ge3$, we have
\begin{equation}
u H_d'(u) + n H_d(u) = (n+1)(n+2)u^2 - 12(n+1)u + 12n(n-1) > 0,
\end{equation}
which holds since the discriminant of the right-hand side is negative for $n\ge3$. Combined with the positive denominator, this implies $dh_d/du>0$ for all $u>0$. Thus each positive charge corresponds to exactly one positive critical radius for $d\ge7$.

\begin{table}[H]
    \centering
    \caption{Discriminant and root structure of $H_d(u)$}
    \label{table2}

    \begin{tabular}{lcccc}
        \hline \hline
         Dimension & Sign of discriminant & Structure of $H_d(u)$ & \\
        \hline
        $d=5$ & $+$  & Two roots, one at $u=0$ \\
        $d=6$ & 0 & Degenerate positive root \\
        $d\ge 7$ & $-$ & No real roots \\
        \hline \hline
    \end{tabular}
\end{table}

These results are summarized in Table~\ref{table2}. Hence, we obtain that $d=6$ is the unique case where charged GB-AdS black holes can admit multiple critical points for a suitable range of charge parameters, while only one single positive critical point exists for $d=5$ and $d\ge7$.

\end{document}